\documentclass[journal,10pt,onecolumn]{IEEEtran}

\usepackage{graphicx}
\usepackage[ruled]{algorithm2e}
\usepackage{algorithmic}
\usepackage{cases}

\begin{document}

\title{Vehicular Edge Computing via Deep Reinforcement Learning}

\author{Qi Qi,~Zhanyu Ma}

%{Shell \MakeLowercase{\textit{et al.}}: Bare Demo of IEEEtran.cls for Journals}

\maketitle

\begin{abstract}
The smart vehicles construct Vehicle of Internet which can execute various intelligent services. Although the computation capability of the vehicle is limited, multi-type of edge computing nodes provide heterogeneous resources for vehicular services. When offloading the complicated service to the vehicular edge computing node, the decision should consider numerous factors. The offloading decision work mostly formulate the decision to a resource scheduling problem with single or multiple objective function and some constraints, and explore customized heuristics algorithms. However, offloading multiple data dependency tasks in a service is a difficult decision, as an optimal solution must understand the resource requirement, the access network, the user mobility, and importantly the data dependency. Inspired by recent advances in machine learning, we propose a knowledge driven (KD) service offloading decision framework for Vehicle of Internet, which provides the optimal policy directly from the environment. We formulate the offloading decision of multi-task in a service as a long-term planning problem, and explores the recent deep reinforcement learning to obtain the optimal solution. It considers the future data dependency of the following tasks when making decision for a current task from the learned offloading knowledge. Moreover, the framework supports the pre-training at the powerful edge computing node and continually online learning when the vehicular service is executed, so that it can adapt the environment changes and learns policy that are sensible in hindsight. The simulation results show that KD service offloading decision converges quickly, adapts to different conditions, and outperforms the greedy offloading decision algorithm.
\end{abstract}

\begin{IEEEkeywords}
vehicle of internet, service offloading decision, multi-task, knowledge driven, deep reinforcement learning
\end{IEEEkeywords}

\IEEEpeerreviewmaketitle

\section{Introduction}
Along with the development of information technology, the vehicles with advanced computation and communication capabilities spring up. With the promising electronic control units (ECUs), the vehicles construct as Vehicle of Internet which can execute multimedia entertainment services, navigation, remote vehicle diagnosis and even AI based automatic driven services. Moreover, unmanned aerial vehicle (UAV) can be used as patrol robot which travels in the area of industry field. These applications bring new experience for the vehicular users. However, the limited capability of the ECU and the unstable of the vehicular network still impacts the quality of the service. Fortunately, edge computing is combined to traditional vehicular network to strengthen its service capability. The edge computing environment providing for vehicles including computation nodes located with Base Station (BS) and many kinds of computational hot points of WLAN (Wireless Local Area Networks) access technology, such as intelligent roadside unit, cloudlet in buildings and so on. All of these computational nodes and the neighboring vehicles can provide service execution for the traveling vehicles. The current vehicular edge computing environment are depicted, which supports the efficient communication, control, and computation requirements of intelligent services for vehicles \cite{1}.

(1)	The edge computing nodes located with multiple BSs always consist of high-performance servers or distributed data centers. The BS edge computing nodes are distributed in different geographic regions, and may be belonged to different communication operators. Accordingly, they may have different serving areas, rental cost and available resources, so as to satisfy the needs of service requirement in the moving vehicles. The vehicles act like thin clients connecting over to the edge computing node through mobile network without the need of accessing to the remote cloud.

(2)	The cloudlet edge computing servers are deployed with wireless APs(Access Points) located at intelligent roadside units, such as the road lamp, intersection, shops or buildings \cite{2} \cite{3}. These nodes provide the vehicles various local capability and services, including computation, networking, storage and applications by limited transmission coverage communication technology. The vehicles within the coverage area can use the powerful computing capabilities to execute their computation intensive services or access some services provided by the roadside cloudlet. This type of nodes in the hierarchy edge computing environment is important, as it is closer to users, has powerful computation potential, and well-connected to the mobile devices via a high-speed local communication technology, which is benefit to provide low-latency computation and rich computational resources.

(3)	The ad hoc vehicular nodes can be also treated as edge computing nodes since the user can exploit the capabilities of the neighboring vehicles to execution some service \cite{4}. The computation for services are offered by neighboring vehicles that have sufficient resources to act as edge computing servers. Consumers need to discover the vehicle edge nodes, be aware of their resources, and communicate and request resources from them. In this scenarios, the resource provider and resource consumer can be transferred according to the service requirement.

Offloading the services to the vehicular edge computing nodes other than hosing the execution on the vehicular device itself, can expand the limited capabilities of the smart vehicles. The above heterogeneous vehicular edge computing nodes have different capability and using scenarios, which provides multiple choices for the vehicular users. The BS edge computing nodes have large scale coverage and high-performance computation. If an application in a fast traveling vehicle requires high computational power, it is more desirable to offload to the BS edge computing node. However, a weakness of the BS edge computing nodes is that the users may experience long latency for data exchange with the corresponding far away nodes through the 3G/4G access network. Long latency would hurt the interactive response, since humans are acutely sensitive to delay and jitter. For the AP edge computing nodes, if too many vehicles access the same wireless network simultaneously during a period, the scarce bandwidth may affect the application QoS. For the real-time applications with real-time constraints, it would be more beneficial for the user to execute the service locally on the ad hoc vehicle node to avoid the long communication time. Since the current vehicular service are complicated and always contain multiple tasks with various requirement, so we should make decision for each task overall considering communication latency, computation capability and vehicular mobility.

Traditionally, the offloading decision algorithms depend on the human experience. The relevant knowledge is abstracted by researchers including the system formulation, algorithm solving and method optimization. However, abstracting knowledge by hand has mainly three limitations. Firstly, the vehicular environment which contains vehicles, edge computing nodes, and access network, cannot has a comprehensive precise, and real-time evaluation. For example, for the vehicular service offloaded to the heterogenous nodes, the task execution progress is always abstracted as Non-Poisson queuing model, which is hard to solve and guarantee the fairness \cite{5}. Moreover, the objective for the offloading decision model is almost non-convex function, and Lagrange Relaxation is need to process the constraint conditions or some specific heuristic algorithms are used to find the near-optimal solutions. Finally, each time the vehicular edge computing environment changing, the offloading decision should be re-computed, which results in more service delay and higher cost. The advantage of edge computing nodes is that they are deployed with the network devices, and can capture multi-dimensionality data from the environment, including vehicular behavior, task information and network status. Therefore, it is necessary that utilizing the data from environment to keep continually learning of task scheduling knowledge and provide offloading policy directly from the environment data.

Deep learning is the hot area for the Computer Vision, Speech Recognition, Natural Language Processing, and even the computer networks, with some remarkable achievements. The deep learning model including DNN (Deep Neural Network), CNN (Convolutional Neural Network), RNN (Recurrent Neural Networks), ResNet, DenseNet and so on \cite{6}. These various types of neural networks utilize the multi-layer network structure and nonlinear transformation, construct the lower-level features, and formulate the abstract and distinguish high-level expression, so that to percept and express of the things.

The service offloading decision in vehicle edge computing environment is affected by many factors, and it can be abstracted to a mapping problem including non-convex optimization with complicated objective function and constraints. The deep learning model divides the complicated mapping problem into several embedded simple mappings depicted by the multiple layers of the model. The iteration optimization based on gradient during the process of training minimizes the loss function which expresses the approximate degree. After the supervised learning process based on the human labeled history data or the solution from heuristic algorithms, the deep learning model can obtain the approximate optimization solution of the complicated mapping problem. When the model is deployed in the real-word setting after the training, the approximate optimization solution of the mapping problem can be independently obtained according to the environment data in real-time.

Considering the environment changing in the future, the deep learning method needs the new labeled data. The service offloading decision for complicated service with different requirement tasks should has the long-term programming and continually learning capabilities. Deep reinforcement learning (DRL) method combines the perceive capability of the deep learning and the decision capability of the reinforcement learning. Reinforcement learning based on the MDP (Markov Decision Process) theory, but it need not to formulate the states transfer probabilities. It depends on the samples of the system states and objective reward from the experience policy, and decides according to the current situation. On the other hand, the effective perceive of deep learning models improves the poor performance for traditional RL methods when the environment contains the high dimension input state space or large action sets.

In this paper, we address the shortcomings of traditional service offloading via a knowledge driven (KD) service offloading decision framework. It includes the decision model with the DRL algorithm to learn the offloading knowledge, and the observation function that is responsible for obtaining the data of vehicular mobility and the edge computing nodes. The KD service offloading decision framework provides a unique platform for various vehicular services selecting the three types of edge computing nodes as they need them. It aims at achieving long-term optimal performance experienced by the vehicular users. The main contributions of this paper are described as follows:

(1) Making the service offloading decision according to the learned long-term optimal decision knowledge for Vehicle of Internet. We propose a DRL based offloading decision model, which understands the resource requirement, the access network, the user mobility.  Importantly, it considers the future data dependency of the following tasks when making decision for a current task from the learned offloading knowledge. By this model, the optimal policy can be obtained directly from the environment without the complicated computation of the offloading solution.

(2) The mobility model of the vehicular edge computing environment is formulated for service offloading decision. Since the vehicle moving impacts the offloading destination selection and it is hard to modeling, we formulate the impact of task delay by mobility according to the accessed vehicular edge computing node. This model can be directly used in the online service offloading learning.

(3) Exploring the A3C algorithm to realize the online optimization offloading decision for the moving vehicles. The offloading decision model is trained for each service in the edge computing node and distributed to the vehicles. The vehicles perform asynchronous online learning when it running the service, and updates the new model to the edge computing node. By the feedback reward in each running time, the KD service offloading decision framework can adapt the environment changing.

In Section II, we first describe the related work about service offloading decision for vehicular edge computing and review import development of joint optimization based on DL and DRL. We then present the architecture of the KD service offloading decision framework and the problem formulation in Section III. The DRL model and corresponding service offloading algorithm are presented with details in Section IV. Performance evaluation are presented in Section V. Section VI concludes the paper and presents the future work.

\section{Related Work}
\subsection{Offloading Decision for Vehicular Edge Computing}
The vehicular edge computing is proposed along with the development of edge computing and VANET technology \cite{1}. The high performance of edge computing  servers provides vehicles more capability for executing complicated application \cite{7}. Qin et al  propose VehiCloud as a service-oriented cloud architecture providing routing service by predicting vehicles future locations. The three edge computing nodes support vehicles offloads their services, such as image detection, speech recognition, web fasten and online game. Generally, complicated service can be divided into multiple tasks which can be executed independently at local ECU, or edge computing nodes. Some part of the tasks contains the dependency by data transmission, which construct as a DAG (Directed Acyclic Graph). Therefore, the offload decision should consider the data transmission between the dependent tasks.

From the mobile cloud computing, there are many valuable works focusing on offloading decision, which includes offload or not, offload volume and offload location, considering service type, user perfect, access technology, network traffic, device capability, edge node property and so on. The offloading decision is extreme complicated, consisting of "single-user to single-node multi-user to single-node and multi-nodes scenarios. The  ingle-user to single-node problem only has one target node and the decision should make a joint optimization for all the tasks in a DAG \cite{8}. If the service arrives in stochastic model, the long-term reward and the cache stability should be considered. The "multi-user to single-node problem focus on the communication interference and resource competition among the users \cite{9} \cite{10}. The service offloading policy should consider the resource allocation in the edge computing nodes and the utilization of the point-to-point cooperation among the users. Finally, the "multi-node problems for single-user and multi-user all pay attention on the selection and cooperation of the edge computing nodes \cite{11} \cite{12} \cite{13} \cite{14}. They analyze the relationship of service offloading volume and the resource of edge computing nodes, and try to avoid to select the hot node and hot line which will result in overload.

The above three problems can be formulated by combinatorial optimization, MDP, Semi- Markov Decision Process, Cooperative Game, or Non-cooperative Game models \cite{11}\cite{14}, with specific objective function and constraints. Since most of these models are NP-hard problems, multi-stage heuristic algorithm, Lagrangian Relaxation Approach, Lyapunov algorithm \cite{10}, Particle Swarm algorithm, Genetic Algorithm, and Simulated Annealing are used to solve the problems in the acceptable time.

\subsection{Combinatorial Optimization Based on Deep Learning}
The deep learning model solves the combinatorial optimization problem by several embedded simple mappings depicted by the multiple layers. After the supervised learning process, it learns the approximate optimization solution by nonlinear fitting. When the model is deployed, the approximate optimization solution can be obtained according to the input data directly.

Some works take use of the deep learning model instead of the traditional formulation and heuristic methods, to find the optimized solution from the high dimensionality data in network and cloud resource management. The Depth Boltzmann Machine is used to the network traffic control and routing \cite{15} \cite{16}. The input traffic matrix is constructed by the number of arriving packets to the router in a period of time. The output of the model is the next-hop router. The labeled data is obtained by the OSPF (Open Shortest Path First) algorithm in each router for training next-hop policy. In additional, the DNN model is used to find the near-optimal solutions of caching placement, user association, and content delivering \cite{17}. This work reduces complexity in the delay-sensitive operation since the computational burden is shifted to the DDN training phase.  The deep learning-based auction mechanism is used for optimization of edge computing resources \cite{18}. The goal is obtaining the max reward of edge computing resource provider by reasonable resource allocation and payment rule, under the condition of incentive compatibility and individual rationality.
\subsection{Combinatorial Optimization based on Deep Reinforcement Learning}
Deep learning method has better generalization ability, without the need to formulate the environment. After the training the deep learning method, it can achieve approximate optimal solution of the optimization results in real time. However, the deep leaning method is trained by labeled data from the human labeling or heuristic algorithms, which can hardly deal with possible changes in the future.

Deep Reinforcement Learning has achieved remarkable success in a range of tasks, from continuous control problems in robotics to playing games like Go and Atari. The DQN (Deep Q-Network) model firstly proposed by DeepMind combines CNN and Q-learning, trains the CNN by the rewards obtained by the Q-learning in each epoch and realizes the directly control from input states to the output policy \cite{19}. Moreover, deep policy gradient expresses the policy by the parameters of the deep neural network, i.e. the probability for choosing each action in each decision epoch. Through the finding of the gradient for the policy, this method can find the approximate optimal policy. A3C (Asynchronous Advantage Actor-Critic) model utilizes two independent deep neural networks, which are taking charge of update policy and providing policy gradient, respectively \cite{20}. The asynchronous gradient descent method is used for optimizing the parameters of the deep neural network.

Taking use of the DRL method to optimizing the resources is different with the playing game, including the input data, action space and the long-term reward related to the optimal objective. The DQN with CNN and Q-learning is used in the cache-enabled interference alignment wireless network \cite{21}, where the system state is channel state and cache stage. The optimal policy is selecting active users for realizing minimization of the throughput. Mao et al. \cite{22} finding the optimal resource management policy by DRL. The resource requirement for all the job arriving in a period is abstracted as an image which is treated as the system state and inputted into the CNN. The resource scheduling decision is made by policy gradient method. Liu et al \cite{23} utilize the LSTM and DNN combined with Q-learning to dynamic optimize energy and resource for VM, respectively. The system state consists job arriving time, resource utilization rate, job status and resource requirement; the action set is the candidate virtual machines for each job. Currently, the A3C algorithm obtain outstanding results for discrete or continuous controlling problems, which is successfully used in video code selection \cite{24}. From the observation of QoE, network bandwidth, residual cache and the supported codes, the A3C algorithm provides a video code selection, which can realize a long-term optimization of user experience. Moreover, A3C method is used in user scheduling and resource allocation in heterogeneous networks \cite{25}, and A3C with the improved training is used in the traffic allocation among multiple paths \cite{5}.

From the above work, DRL has two advantages, i.e. adaptively and long-term planning. Comparing with the heuristic algorithms, the DRL can adjust the policy according to the environment changing and find the temporarily suboptimal decision for the long-term optimal solution.
\section{System Architecture and Problem Formulation}
The KD service offloading decision framework for vehicular edge computing, provides a unique platform for various services and all the controlled vehicles. It includes the decision model with the DRL algorithm to learn the service offloading knowledge, and the observation function that is responsible for obtaining the environment data of vehicular mobility and the edge computing nodes. Since the services have various numbers of tasks, the KD service offloading decision framework keeps one basic decision model for each service. The basic decision model is trained at the powerful edge computing nodes, such as BS edge computing nodes, and then distributed to the vehicles for the real-world service offloading decision. As the DRL model contains a reward for each decision, the model can be trained online continually by the running services. During this procedure, the parameters are transmitted from the vehicles to the BS edge computing node for updating the basic model periodically.

The number of accessible edge computing nodes for the vehicles are different, and it may change when the vehicle moving. The observation function sorts the accessible nodes for the three types of the edge computing nodes according to their computation power, and adopts the fixed number of each type of the nodes as the candidate offloading destination.

The heterogeneous resources provided by vehicular edge computing nodes are abstracted to several services with specific functions and parameters. The multi-task in a complicated application is modeled as a specific dataflow graph and denoted by DAG, depicted in Fig. 1. Due to the differences existing in the edge computing nodes, including several cloudlets, BS that contains computing and storage capability and the neighboring vehicles on the road.

\begin{figure}[!t]
\centering
\includegraphics [width=3.5in]{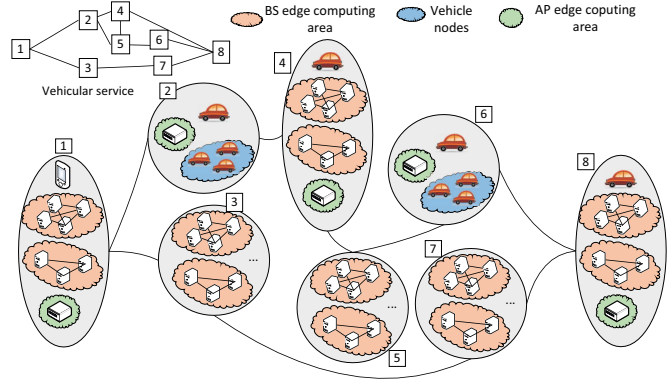}
\caption{The vehicular service denoted by a DAG.}
\label{Fig.2}
\end{figure}

\begin{table}[!t]
\renewcommand{\arraystretch}{1.3}
\caption{The Notations Defination}
\label{table_example}
\centering
\begin{tabular}{c|l}
\hline
\bfseries Notations & \bfseries Defination\\
\hline
$f_i^t$ & The number computation commands of task i \\
$d_{i-1,i}$ & The data volume from task $i$-1 to current task $i$ \\
$d_i^u$ & The interactive data volume between the local vehicle \\
             &and the task execution node $i$\\
$f_{c_i} $ & The CPU frequency of  edge computing node $c_i$\\
$b_{c_i}$ & The link bandwidth between the vehicle to node $c_i$ \\
$b_{c_{i-1},c_i }$ & The link bandwidth between the nodes of \\
                            &task $i$ and task $i-1$ \\
$D_i $ & The execution time of task $i$\\
$w$ & The channel bandwidth of BS edge computing node \\
$q_v $ & The transmission power of the vehicle $v$\\
$g_{v,c_i}$ & The channel gain between the vehicle $v$ and BS $c_i$\\
$\varpi_0$ & The background noise power of the BS \\
$W_{min}$ & The minimum contention window\\
$m_b$ & The maximum back off stage\\
$p_c$ & The probability of a collision seen by a transmitted packet\\
$p_t$ & The probability of a device transmitting a packet \\
          &in a slot of time\\
$\tau_s$ & The average channel busy time due to packet transmitting \\
$\tau_c$ & The average channel busy time due to collision\\
$L$ & The average packet payload size of vehicular service \\
$h$ & The number of devices connect to a AP contending\\
       & for transmitting data at the same time\\
$\mu_i^{-1}$ & The mean time of the task $i$ execution\\
$\eta_{c_i}^{-1}$ & The mean residence time in one access area of the node\\
$N_{c_i}^h$ & The number of handoff times during task $i$ execution\\
$z^*$ & The largest transmission range between to vehicles\\
$z_{min}$ & The minimum inter-vehicle distances\\
$z_{max}$ & The maximum inter-vehicle distances\\
$p,q,\beta$ & The vehicle density parameters\\
$R_{c_i}$ & The resource usability of the vehicular node $c_i$ for task $i$\\
$D_h$ & The migration time when a vehicle changes\\
           & edge computing node\\
\hline
\end{tabular}
\end{table}

\subsection{Task Execution Delay}
The delay for each task offloaded to outside the vehicle contains four parts. The computation requirement for the task is $f_i^t$, which can be denoted by the number of commands. The task execution time is calculated by the total number of CPU cycles required to accomplish the computation task $i$.  The parameter $d_{i-1,i}$ is the data volume transferred from precursor tasks, and $d_i^u$ is the interactive data volume between the vehicle and the application during task $i$ execution. The data transfer delay for transmitting the data from the forward task $i-1$ to the task $i$, which is calculated by input data volume $d_{i-1,i}$ and the bandwidth of the link between the two execution nodes of task $i$ and task $i-1$, $b_{c_{i-1},c_i }$. The interactive delay is the time of transmitting the data from the vehicle to the offload destination, which is calculated by interactive data volume $d_i^u$ and the wireless bandwidth of the access network link $b_{c_i}$. Thus, the task $i$ execution time in edge computing node $c_i$ is:
\begin{equation}
\label{eqn_example}
D_i=\frac{f_i^t}{f^{c_i}} +\frac{d_i^u}{b_{c_i}}+\frac{d_{i-1,i}}{b_{c_{i-1},c_i}}
\end{equation}
Moreover, for analyzing the bandwidth of the offloading destination, we consider two cases, the vehicle accessing the edge computing node at a BS and the vehicle accessing the edge computing node at an AP or the vehicle accessing another vehicle by WLAN.

The edge computing nodes which are deployed with the BS can be accessed through the 4G technology. If the vehicular user chooses the BS edge computing node $c_i$ to execution task $i$, the uplink communications are serving multiple users. Assume the channel bandwidth is $w$, the set of accessed devices in the BS is $U_{c_i}$, the transmission power of the vehicle $v$ is $q_v$, the channel gain between the vehicle $v$ and BS $c_i$ is $g_{v,c_i}$, and the background noise power is  $\varpi_0$. According to \cite{26}, the data transmission rate between the edge computing node and the vehicle $v$ is:
\begin{equation}
\label{eqn_example}
b_{c_i}=w\log_2{\frac{1+q_vg_{v,c_i}}{\varpi_0+\sum_{u\in{U_{c_i}},u\neq{v}}{q_ug_{u,c_i }}}}, c_i\in{BS}
\end{equation}

The edge computing nodes which are deployed with the WLAN AP has limited transmission coverage and a stochastic characteristic for communicating such that a vehicle can offload their tasks only when it stays within the coverage area for at least a certain amount of time with the mobile vehicle going through its coverage region due to limited communication capacity and time-varying wireless channels. The channel model is important since that the data rates of WLAN changes with different network conditions, which cannot be simply classified into "good" or "bad" states. Assume $h$ devices are contending for transmitting data at the same time. $W_{min}$ is the minimum contention window and $m_b$ is the maximum back off stage. The parameter $p_c$ is the probability of a collision seen by a transmitted packet. The probability that a device transmits a packet in a slot of time is deduced as:
\begin{equation}
\label{eqn_example}
p_t=\frac{2(1-2p_c)}{(1-2p_c )(W_{min}+1)+p_cW_{min}[1-(2p_c)^{m_b }]}
\end{equation}

Assume $\tau_s$ is the average channel busy time due to a successful transmission, and $\tau_c$  is the average busy time when the channel is suffering from a collision. $L$ is the average packet payload size. According to \cite{27}, the data rate of the edge computing node with WLAN access can be calculated as follow:
\begin{eqnarray}
\label{eqn_example}
b_{c_i}&=&\frac{h p_t L}{(1-p_t)(1+\tau_s)+[(1-p_t)^{1-h}-(1-p_t-h p_t)]p_t }, \nonumber \\
        && c_i\in{AP}
\end{eqnarray}

\subsection{Performance of Edge Computing Node}
The performance of the resource provided by the tree types of edge computing node is formulated. The vector $C=(c_1,c_2,\cdots,c_m)$ denotes the $m$ available edge computing nodes surrounding the vehicle, including the neighboring running vehicles, the accessed BS with computing capability and various the roadside units, where the task can be offload. The parameter $f_{c_i}$ is the computation capability (i.e., CPU cycles per second) of the execution node $c_i$ when the task is offloaded. When calculating the data transmission delay, the link of the edge node $c_i$  includes two types: the link from the vehicle to an edge computing node which can be calculated by (2) or (4), and the link between two non-mobility edge computing nodes, which can be measured.

As the mobility model is different in cases of vehicle accessing the BS or AP nodes and vehicle accessing the neighboring vehicle nodes, we analyze the impact on task execution of the cases, respectively.

(1) BS or AP edge computing nodes. The vehicle may access to multiple edge computing nodes geographical distributed. The edge computing node deployed with BSs and APs usually provides resource to the vehicles that are connected, and the offloaded task may be migrated to a new node according to the vehicle moving. Therefore, considering the node geographical location, the number of traversed access networks during the vehicle movement impacts the service delay. Assume the execution time of task $i$ is exponentially distributed with mean value of $\mu_i^{-1}$. The mean residence time in one access area of the node $\eta_{c_i}^{-1}$ is the general continuous random variable with the probability density function of $f_{res}(x)$. Assume $f_{res}^*(x)=\int_{0}^{\infty}f(x)e^{-\lambda x}dx$ is the Laplace-Stieltjes Transform for the $f_{res}(x)$, and then $\eta_{c_i}^{-1}=\int_{0}^{\infty}xf(x)dx$. The number of handoff times during task $i$ execution is $N_{c_i}^h$ that can be deduced according to \cite{28}
\begin{equation}
\label{eqn_example}
N_{c_i}^h=\sum_{k=0}^{\infty}kP(k)=\sum_{k=1}^{\infty}\frac{k\mu_{i}}{\eta_{c_i}}[1-f_{res}^*(\mu_i )]^2 [f_{res}^* {\mu_i }]^{k-1}=\frac {\eta_{c_i}}{\mu_i}
\end{equation}

(2) Neighboring vehicular nodes. The availability of resources provided by the neighboring vehicles construct as an ad hoc cloud environment. The vehicle node usability is determined by the communication link when all the vehicles are moving on the highway. The inter-vehicle distance is called distance headways. We consider a multi-lane highway, and the vehicle density is time-variant.  Assume the transmission range of the vehicles is $z^*$. When the distance headway of any two vehicles becomes larger than $z^*$, the communication link is failed, and the offloaded task cannot be used. To model the variation of the distance headway, a discrete-time finite-state Markov chain is used according to \cite{29}.

Let $X=\{x_1, x_2, x_3, \cdots, x\}$ be the distance headways between a vehicle and one of its neighbors. The random variables $x_i \in [z_{min},z_{max}]$ are the distance headways in each time step during the task execution. Here, $z_{min}$ and $z_{max}$ are the minimum and maximum inter-vehicle distances, respectively. Let  $z$ be the unit length of distance headway changing. The state $X_j$ can be the ranges of a distance headway between $[x_j, x_j + z]$ at a time step. The state $X_{max}$ corresponding to the distance headway just larger than $z_{max}$. Then, let $x_j=z_{min}+j z$. Within a time slot, a distance headway in state $X_j$ can transit to the next state, the previous state, or remain in the same state with probabilities $p_j$, $q_j$, or $l_j$, respectively. $X_{z^*}$ state corresponding to the distance headway with the largest communication range $z^*$. Assume the time step is $k$ seconds. During a task execution time with mean value of $u_i^{-1}$ , the period is divided into $1/k\mu_i$ time steps. If the largest distance headways during the task execution is smaller than $z^*$, the node usability is 100\%. For example, when the vehicles density is large in the traffic jam, the distance headways may be always less than $z^*$. Otherwise, the node usability is the sum of the state probability when the state number is less than $X_{z^*}$.

According to \cite{29}, the state transition probability is as the follows.
\begin{eqnarray}
p_j &=&p[1-\beta(1-\frac{z_{min}+j z}{z_{max}})]  \\
q_j &=&q[1-\beta(1-\frac{(z_{min}+j z}{z_{max}})] \\
l_j  &=&1-p_j-q_j, 0 \leq p,q,\beta \leq1
\end{eqnarray}
The parameters $p$, $q$ and $\beta$ can be set in terms of the vehicle density. Generally, the value of $\beta$ increases as the vehicle density increases, and thus the dependency on the state value increases. When the vehicle density is at a low level, $\beta$ is close to zero, and the transition probabilities are independent of the state. We can obtain the one-step transmission matrix:
\begin{equation}
Q = \left(\begin{IEEEeqnarraybox*}[][c]{,c/c/c/c/c/c,}
l_0&p_0&0&\cdots&\cdots&0\\
q_0&l_1&p_1&\cdots&\cdots&0\\
\cdots&\cdots&\cdots&\cdots&\cdots&0\\
0&0&0&\cdots&l_{z^*}&0\\
0&0&0&\cdots&\cdots&0
\end{IEEEeqnarraybox*}\right)
\end{equation}
Let the $P_j^I$ denotes the probability of the state $X_j$ in the $Ith$ step state transmission. The initiation state probability vector is $\pi(0)=(P_0^0,P_1^0,P_2^0,\cdots,P_{X_{z^*}}^0,\cdots, P_{X_{max}}^0)$. Assume in the time step $\zeta$, $0 < \zeta \leq 1/k\mu_i$, $\pi(\zeta)=(P_0^\zeta,P_1^\zeta,P_2^\zeta,\cdots,P_{X_{z^*}}^\zeta,\cdots,P_{X_{max}}^\zeta)$ is the probability vector of the states. From the property of Markov chain,
\begin{equation}
\label{eqn_example}
\pi(\zeta)=\pi(0) Q^{\zeta}=\pi(0)\times \underbrace {(Q\times \cdots \times Q)}_{\zeta}
\end{equation}
Then, the usability of vehicular node $c_i$ for task $i$ is
\begin{equation}
R_{c_i}=\sum_{j=0}^{z^*}P_j^\zeta
\end{equation}

If the task is offloaded to edge computing node deployed with BS or AP, the task execution delay should be added the handoff delay, assumed as $D_h$. If the task is offloaded to another vehicle node (VN) and the distance of the two vehicles are larger than $z^*$, the task may fail, and the task execution delay should be added a re-execution delay at the local vehicle, i.e. $c_i=lo$
 \begin{numcases}{D_i^\prime=}
{D_i+{D_h}{N_{c_i}^h},}                                                                             &{}{$c_i \in {BS \cup AP} $}\nonumber \\
{D_i+R_{c_i}(\frac{f_i^t}{f^{c_i=lo}}+\frac{d_{i-1,i}}{b_{i-1,c_i=lo}})}, &{}{$c_i\in {VN}$}
\end{numcases}

It is generally known that four basic topologies exist in an application, sequence, parallel, selective and loop. These topologies are able to construct the vast majority of composited applications. Assume $M$ tasks are contained in a vehicular service. When the task $i$ is a parallel task, $F(i)$ is dedicated that whether the task $i$ is the longest one.
\begin{equation}
\label{eqn_example}
\min D_s=\min \sum_{i=1}^M F(i)D_i^{\prime}
\end{equation}

\section{Knowledge Driven Service Offloading Decision}
DRL extends the well-known traditional reinforcement learning to enable end-to-end system control based on high-dimensional sensory inputs. Different from supervised learning, reinforcement learning does not learn from samples provided by an experienced external supervisor. Instead, it has to operate based on its own experience despite that it faces with significant uncertainty about the environment. In this section, we mainly focus on the offloading decision of the above framework and propose the KD service offloading decision to reduce execution time for running vehicular service by an A3C based algorithm.
\subsection{Deep Reinforcement Learning Model}
A DRL model comprises an agent that interact with the environment based on its observations. At each time step $t$, the environment is in the state $s_t$, and the agent executes an action $a_t$. Then, the environment may transfer to any achievable next stage $s_{t+1}$ by some probability, and the agent receives a reward $r_{t+1}$. The long-term goal of the agent is to maximize the cumulative reward it earns, by taking a policy $\pi$ which adapts its action according to its observations. The accumulated return for the step $t$ with a discount factor $\gamma$ of the future rewards is $R^t=\sum_{k=0}^{\infty}{\gamma^k r_{t+k}}$.  From the goal of the agent, the value of state $s_t$ is the expected return for following the policy $\pi$, which is defined as $V^\pi(s_t)=E[R^{t}\vert s=s_t]$.  When taking use of the deep neural network as the function approximate, its parameters is denoted as $\theta$. Currently, there are various RL algorithms for updating $\theta$, including DQN,  Deep Deterministic Policy Gradient (DDPG) and A3C.

The Actor-Critic architecture derives from the policy-based model-free methods, which combines the advantage of Q-learning and policy gradients. It directly parameterizes the policy, denoted by $\pi(a_t \vert s_t;\theta)$ and update the parameter $\theta$ by calculating the gradient ascent on the variance of the expected accumulated return $R^t$ and the learned value function under the policy $\pi$, i.e. $R^t-V^\pi(s_t )$.  In the Actor-Critic architecture, the actor is the policy function $\pi(a_t \vert s_t;\theta)$, which decides the action currently based on learning the policy that can achieve the highest reward. Then the environment changes, and the agent receives corresponding reward. While, the critic uses this reward to evaluate the current policy according to the TD-error between current reward and the estimation of the value function $V(s_t;\theta_v )$. The value function can be calculated by value-based learning method. For the DRL model, the policy function and the value function are both neural networks. The feedback of TD-error is used to update the actor network parameter $\theta$ for adding the probability of selecting the better action as well as the critic network parameters $\theta_v$ for obtaining more accurate estimation value. By this way, the AC algorithm learns the policy function and value function together. Along with the iteration, the critic archives more accurate estimation and the actor makes better selection and finally, the system converges.

The A3C algorithm based on AC algorithm utilizes asynchronous multi-threads as multiple actors to train DNN reliably. The multiple actor learners running in parallel can explore different parts of the environment with different policies. By this way, these updates of parameters are less correlated in time than a single agent, so that the replay memory of traditional DQN do not needed. Moreover, the training time is reduced.

Each agent keeps its own parameters of the actor network and the critic network, and also a global actor network and a global critic network. Each time the agent updates its parameters of the two networks, then it submits the parameters to the global networks. When the global network receives the new parameters, it transmits them to the agents. After several iterations, the two networks converge.

The A3C algorithm uses $k$ steps rewards to update the parameters. The critic network updates the parameters $\theta_v$ of the value function $V(s_t;\theta_v)$  and make the value function to approximate the real reward \cite{20}.
\begin{eqnarray}
&&\hat{G}(s_t )=r_t+\gamma r_{t+1}+\gamma ^2 r_{t+2}+\cdots+\gamma^{k-1} r_{t+k-1 }\nonumber\\
                   &&+\gamma^k V(s_{t+k};\theta_v)=\sum_{i=0}^{k-1}\gamma^i r_{t+i}+\gamma^k V(s_{t+k};\theta_v)
\end{eqnarray}
The A3C algorithm follows the AC model and defines the advantage function as the estimation of the difference of the real reward and the value function that is the estimation at the state $s_t$ under the parameters of $\theta_v$.
\begin{equation}
A(s_t,a_k;\theta,\theta_v )=\hat{G}(s_t )-V(s_t;\theta_v )
\end{equation}
When update the function value, we try to minimize the difference, i.e.
\begin{equation}
\min_{\theta_v}[\hat {G}(s_t)-V(s_t;\theta_v)]^2
\end{equation}
Then the critic network update performed by the gradient
\begin{equation}
d\theta_v \leftarrow d\theta_v+\frac{\partial {A(s_t,a_k;\theta^{\prime} ,\theta^{\prime} _v)^2}}{\partial{\theta^{\prime}_v} }
\end{equation}
The actor network update performed by the gradient
\begin{eqnarray}
&d\theta \leftarrow &d\theta+\nabla_{\theta^{\prime}}\log \pi(a_t \ | s_t;\theta^\prime)A(s_t,a_k;\theta^{\prime},\theta^{\prime}_v) \nonumber\\
                              &&+\delta \nabla_{\theta^\prime}H (\pi{s_t;\theta^{\prime}})
\end{eqnarray}
Here, $H$ is the entropy. The A3C algorithm uses the entropy of the policy $\pi$ to the objective function of network update, which prevents the premature convergence to suboptimal deterministic policy. The hyperparameter $\delta$ controls the strength of the entropy regularization term \cite{20}.

\subsection{A3C Based Vehicular Service Offload Algorithm}
We enable the DRL model to learn the long-term optimal service offloading knowledge, which is the mapping of the environment observation to the offloading destination, represented by a deep neural network. The observation includes the performance of the edge computing nodes, the status of the vehicular nodes, the vehicle moving speed and the task requirements. The neural network provides an expressive and scalable way to incorporate a rich variety of observations into the service offloading policy. We consider the offloading decision function in the vehicle act as the agent of the DRL model, which interacts with the vehicular edge computing environment through a sequence of observations, actions and rewards. The goal is to select actions in a fashion that maximizes cumulative future reward for all of the tasks in a service.

The state space of the vehicular edge computing is defined as follows: $S=(T,C,v)$. $T$ is a tensor of the task profile, which is defined as $T=(f_i^t, d_i^u, d_{i-1,i})$.  $C$ is the tensor that expresses the current state of the offloading destination nodes including the edge computing nodes and vehicular nodes, $C= (c_{type}, f_{c_i}, b_{c_i}, N_{c_i}^h , R_{c_i})$. The parameter $c_{type}$ is the node type including BS edge computing node, AP edge computing node, and vehicle node. The $v$ is the moving speed of the vehicle. The action space is $A=(a_{local},a_1,\cdots,a_m,\cdots)$, denoting that a task is executed locally on the vehicle, i.e., $a_{local}=1$, or offload to the accessible edge computing nodes $n$ (i.e., $a_m=1$).

Since the number of accessible edge computing nodes for a vehicle is larger than the action space, we choose fixed number of the three types of nodes according to their computation power and construct the action set. Otherwise, if the number of nodes is less than the length of the actions set, we set some pseudo nodes with poor performance to fill the action space. The reward for each decision slot $r_t$ is determined by task execution delay in real set, i.e. $D_i^{\prime}$. The vehicular mobility may result in the task migration or node unavailable. Additionally, when there exist the parallel tasks, only the task with longest delay is considered in the learning process. The goal of the service offloading is achieving the minimized expected cumulative discounted reward for all of the $M$ tasks.
\begin{equation}
J=E[1/M \sum_{t=1}^M \gamma^{M-1} r_t]
\end{equation}

The A3C based vehicular service offloading algorithm is depicted as follows. The DRL model is firstly training at the powerful edge computing node by the generated service data. Then, when a vehicle initiates a service, the corresponding pre-trained model is fetched, and the online learning by the actor network and the critic network is performed. The vehicles running the same service can update the parameters of the DRL asynchronously. Finally, the offloading decision knowledge can be online learned by vehicle, which can adapt the environment changing.
\begin{algorithm}
\algsetup{linenosize=\tiny} \scriptsize
\caption{The vehicular service offloading decision algorithm}
\LinesNumbered
\uIf{training at the powerful edge computing node}{
Initialize the critic network with parameter $\theta_v$ \\
Initialize the actor network with parameter $\theta$\\
Obtain the threads of the CPU\\
}\Else{Fetch the corresponding model for a service from edge computing node \\
}
\While{DataSet is not empty or All vehicles running the service}{
Set the gradient of the two networks $d\theta=0$ and $d \theta_v=0$ \\
Synchronous thread parameters by global parameters $\theta^\prime=\theta, {\theta_v}^\prime=\theta_v$ \\
Obtain the vehicular edge computing nodes status and the task profile \\
Construct the environment state $S_t$\\
\For{$t=1, t \leq M$}{
  Select the task offload location $a_t$ according to policy $\pi(a_t|s_t;\theta^\prime)$ in actor network\\
  Calculate the reward $r_t$ and construct the new environment state $S_{t+1}$ \\
  t=t+1\\
}
  Obtain the $R=V(s_t,\theta_v^\prime)$ from the critic network\\
\For{$t=M, t \geq 1$}{
  $R=r_t+\gamma R$ \\
  Calculate the accumulate gradient $d\theta_v$ for critic network by (17)\\
  Calculate the accumulate gradient $d\theta$ for actor network by (18) \\
  t=t-1\\
}
\uIf{training at the BS edge computing node}
{Asynchronous update $\theta_v$ and $\theta$, respectively}
\Else
{Send the $d\theta_v$  and $d\theta$ to BS edge computing node}
}
\end{algorithm}

\section{Performance Evaluation}
In this section, the performance of the KD service offloading decision is evaluated. First, we introduce the simulation scenarios of the vehicular edge computing environment and the vehicular applications used to train the DRL model. Second, we evaluate the usability of the edge computing nodes based on the proposed mobility model with difference parameters. Then, the DRL model for offloading decision is analyzed with different parameters. Finally, the service offload decision policy based on the A3C algorithm is evaluated by comparing with greed algorithm.
\subsection{Simulation Scenarios}
In the simulation, we generate various service scenarios that include different number of tasks and the three types of edge computing nodes for training the DLR model. The vehicular environment contains two BS edge computing nodes, with frequency of 560 and 676 , two AP edge computing nodes with frequency of 526 and 430, and six accessible neighboring vehicular nodes with frequency of 124, 120, 177, 144, 165 and 130. When each task execution, the computation capabilities of theses nodes changes according to a normal distribution with standard deviation as 5. The bandwidths between BS to BS, BS to AP, and AP to AP, AP to vehicle are set to be 100 M. The bandwidth of BS to vehicle is set to be 50 M, while vehicle to vehicle is set to be 300 M.

We design several complicated services for vehicular users which contains multiple tasks deployed by in Docker. For example, the location sight recognition service for vehicle drivers includes image pre-process, image segment,  object detection \cite{30}, image classification \cite{33}, image auto-annotation, voice translation, image-voice synthesizing and recommendation based on location \cite{35}.  We generate the dataset according to these services including group of tasks for each service. The service data contains the required CPU cycles, user transmitting data volume and dependency data volume for the previous task. Moreover, two or three tasks can be emerged into one task in the test, so that they may be deployed in one node. Therefore, in the following test, the task number may change from 5 to 30 according to the setting.

\subsection{Usability of Offload Impacted by Vehicular Mobility}
We generate user's moving parameters according to the mobility model introduced in Section IV. We first evaluate the impact on the service delay of BS edge computing nodes and AP edge computing nodes.  When the resident time $\eta_{c_i}^{-1}$ increases, the handoff times decrease; but when the mean execution time  of task i ${\mu_i^{-1}}$  increases, the handoff times rise. Therefore, if the vehicle keeps a high moving speed and the task execution time is long, the vehicular user may suffer a task migration, which adds the service delay.

Moreover, according to the model of vehicle mobility in Section III. B, we analyze the vehicle node usability impacted by parameters of $\beta$, $Z^*$ and the $p$. The $Z^*$ is set to be the 30th states, with the largest states of the distance headway as 40. We can see when the vehicular density parameter $\beta$ increases, the node usability increases. When the value of $\beta$ is close to zero, the vehicle density is at a low level, and the transition probabilities are independent of the state. Since high density results in the large dependency of the mobility transfer state, the node usability may be bigger when the $p$ and $q$ with little higher values. Generally, the parameter $\beta$ does not impact the node usability extremely. In the vehicle mobility model, the parameter $p$ indicates the probability of the distance headway increasing. When the vehicle speed is fast, the headway may increase, which results in the higher probability of the neighboring node inaccessible. Then, the usability of the vehicular node decreases with the rising of parameter $p$. When the largest accessible distance headway increases, i.e. $Z^*$, the usability of the vehicular node increases linearly in the mobility model. Then in the following analysis, we use the different parameter combination to generate the vehicular nodes.

\subsection{Analysis of the DRL model}
We analyze the different parameters of DRL model for understanding the convergence and gains of the KD offloading algorithm. Firstly, we use the five layers DNN as both of the actor network and the critic network for learning the data distribution of the task offloading. The neural cell of the input layer is the dimension number of environment state, the second layer and the third layer contain 64 neural cells. The output of the critic network is one neural cell which express the value of the input state. The output layer of the actor network uses the softmax function to obtain the probability of the actions under the policy. We use the ReLU (Recitified Linear Unit) as the activation function. The model is training on the PC server with 2.3GHz 4 cores CPU and 8 G memory, which provide 4 threads for the asynchronous learning.

We generate the services containing 10 serial tasks with data dependency.  The CPU requirement of the 10 tasks are as follows: four tasks need 5K commands, 3 tasks need 2K commands, and 3 tasks need 9K commands. Moreover, for generating diversified services,  the requirements follows a normal distribution with standard deviation as 500. The data transmitted between the tasks are set to be 2G and follows a normal distribution with mean of 500. The results of KD service offloading decision model in 80000 service samples are depicted in At the same time, the average task delay of the service converges at the long-term optimal.

We compare the average task delay by different entropy hyperparameter $\delta$ value in the DRL model. From the results, choosing $\delta$ as 0.01 can archive the best result, which can prevent the premature convergence to suboptimal policy by appropriate strength of exploration. The average task delay decreases along with the discount factor $\gamma$. For the complicated vehicular service, we choose the $\gamma$ with large value which means the more future execution tasks performance is considered when selecting the current offloading destination. However, when the $\gamma$ value increases, the model should take more service samples to train for convergence. For example, when we choose $\gamma=0.99$, 60000 samples are generated to train.

\subsection{Service Delay Comparison}
We compare the KD offloading decision algorithm with the greedy offloading decision algorithm. The greedy offloading decision algorithm always selects the optimal  destination, i.e. the node with least delay for each task in a service. We measure the average delay for all tasks achieved by the greedy offloading decision algorithm and the KD service offloading decision after the training.

\section{Conclusion}
This paper targets the problem of service offloading decision in vehicular edge computing environment. Our work goes beyond existing approaches by considering knowledge-driven method to obtain the optimal offloading policy. The current vehicular services always contain multiple tasks with data dependency. We propose a KD service offloading framework by exploring the DRL model to find the long-term optimal service offloading policy. Focusing on the types of vehicular edge computing nodes, the service delay as learning reward is formulated which contains the factors of node accessible due to the vehicle mobility. The offloading decision model for each service is trained at the powerful edge computing nodes, and then distributed to the vehicle. The vehicles perform asynchronous online learning when it running the service, and updates the new model to the edge computing node. By the feedback reward in each running time, the online learning KD service offloading decision framework can adapt the environment changing. We generate several datasets for training the KD service offloading decision. The advantage of the KD service offloading decision is obviously when there are more data dependency between tasks in a service.

\ifCLASSOPTIONcaptionsoff
  \newpage
\fi


\begin{thebibliography}{1}
% You can use other form of bib file by changing here...

\bibitem{1}
 X.~Hou, Y.~Li, M.~Chen, D.~Wu, D.~Jin, S.~Chen, ``Vehicular Fog Computing: A viewpoint of Vehicles as the Infrastructure,"\emph{IEEE Trans. Vehicular Tech.}, vol. 65, no. 6, pp. 3860 - 3873, Jun. 2016




\bibitem{2}
R.~Kim, H.~Lim, B.~Krishnamachari, ``Prefetching-Based Data Dissemination in Vehicular Cloud Systems," \emph{IEEE Trans. Vehicular Tech.}, vol. 65, no. 1, pp. 292-306, Jan. 2016


\bibitem{3}
C.~Wang ; Y.~Li ; D.~Jin ; S.~Chen, ``On the Serviceability of Mobile Vehicular Cloudlets in a Large-Scale Urban Environment", \emph{IEEE Trans. Intelligent Transportation Systems.}, vol.17, no. 10, pp. 2960 - 2970, Oct. 2016
\bibitem{4}
X. Chen, L. Jiao, W. Li, X. Fu, ``Efficient Multi-User Computation Offloading for Mobile-Edge Cloud Computing," \emph{IEEE/ACM Trans. Netw.}, vol. 24, no. 5, pp. 2795 - 2808, Oct. 2016
\bibitem{5}
Z.~Xu, J.~Tang, J.~Meng, W.~Zhang, Y.~Wang, C.~H.~Liu, D.~Yang, ``Experience-driven Networking: A Deep Reinforcement Learning based Approach", \emph{IEEE International Conference on Computer Communications (INFOCOM)}, 15-19 April 2018.

\bibitem{7}
K.~Zhang, Y.~Mao, S.~Leng, Y.~He, Y.~Zhang, ``Mobile-Edge Computing for Vehicular Networks: A Promising Network Paradigm with Predictive Off-Loading,"\emph{IEEE Vehicular Technology Magazine}, vol. 12, no. 2. pp. 36-44, Apr. 2017
\bibitem{8}
Y.~Mao, J.~Zhang, K.~B.~Letaief, ``Dynamic Computation Offloading for Mobile-Edge Computing with Energy Harvesting Devices  \emph{IEEE J. Sel. Areas in Commun.}, vol. 34, no. 12, pp. 3590-3605, Dec. 2016
\bibitem{9}
X.~Chen, L.~Jiao, W.~Li, X.~Fu, ``Efficient Multi-User Computation Offloading for Mobile-Edge Cloud Computing  \emph{IEEE/ACM Trans. Netw.}, vol. 24, no. 5, pp. 2795-2808, Oct. 2016
\bibitem{10}
X.~Lyu, W.~Ni, H.~Tian, et al, ``Optimal Schedule of Mobile Edge Computing for Internet of Things Using Partial Information  \emph{IEEE J. Sel. Areas in Commun.}, vol. 25, no. 11, pp. 2606-261, Nov. 2017

\bibitem{11}
Y.~Kim, J.~Kwak, S.~Chong,``Dual-side Optimization for Cost-Delay Tradeoff in Mobile Edge Computing,\emph{IEEE Trans. Vehicular Tech.}, vol. 67, no. 2, pp. 1765-1781, Feb. 2018.
\bibitem{12}
Y.~Zhou, F.~R.~Yu, J.~Chen, Y..~Kuo, ``Resource Allocation for Information-Centric Virtualized Heterogeneous Networks with In-Network Caching and Mobile Edge Computing," \emph{IEEE Trans. on Vehicular Tech.}, vol. 66, no. 12, Dec. 2017, pp. 11339-11351



\bibitem{13}
Y.~Wang, M.~Sheng, X.~Wang, L.~Wang, J.~Li, ``Mobile-Edge Computing: Partial Computation Offloading Using Dynamic Voltage Scaling," \emph{IEEE Trans. on Commun.}, vol. 64, no. 10, pp. 4268-4282, Oct. 2016.
\bibitem{14}
H.~Cao, J.~Cai, ``Distributed Multiuser Computation Offloading for Cloudlet-Based Mobile Cloud Computing: A Game-Theoretic Machine Learning Approach,"\emph{IEEE Trans. Vehicular Tech.}, vol. 67, no. 1,
pp. 752 - 764, Jan. 2018
\bibitem{15}
N.~Kato, Zubair Md. Fadlullah,et al., ``The Deep Learning Vision for the Heterogeneous Network Traffic Control Proposal, Challenges, and Future perspective \emph{IEEE Wirel. Commun.}, vol. 24, no. 3, pp. 146-153, Jun. 2017.

\bibitem{16}
B.~Mao, Z.~M.~Fadlullah, F.~Tang, N.~Kato, et al., ``Routing or Computing? The Paradigm Shift Towards Intelligent Computer Network Packet Transmission Based on Deep Learning," \emph{IEEE Trans. on Computer}, 66 (11), 2017, pp. 1946-1960
\bibitem{17}
L.~Lei, L.~You, G.~Dai, T.~Xuan Vu, D.~Yuan, S.~Chatzinotas, ``A Deep Learning Approach for Optimizing Content Delivering in Cache-Enabled HetNet," \emph{International Symposium on Wireless Communication Systems (ISWCS)}, Aug. 28-31, 2017, pp. 449-453.
\bibitem{18}
N.~C.~Luong, Z.~Xiong, P.~Wang, D.~Niyato, ``Optimal auction for edge computing resource management in mobile blockchain networks: A deep learning approach," \emph{IEEE International Conference on Communications (ICC)}, May 20-24 2018.
\bibitem{19}
V.~Mnih, K.~Kavukcuoglu, D.~Silver, ``Human-level control through deep reinforcement learning," \emph{Nature}, vol. 518, pp. 529-534, Feb. 2015.
\bibitem{20}
V.~Mnih, A.~P.~Badia, M.~Mirza, et al. "Asynchronous methods for deep reinforcement learning,�\emph{International Conference on Machine Learning (ICML)}, Jun. 19-24 2016.
\bibitem{21}
Y.~He, Z.~Zhang, F.~R.~Yu, et al., ``Deep-Reinforcement-Learning-Based Optimization for Cache-Enabled Opportunistic Interference Alignment Wireless Networks� \emph{IEEE Trans. on Vehicular Technology}, vol. 66, no. 11, pp. 10433-10444, Nov. 2017.
\bibitem{22}
H.~Mao, M.~Alizadeh, I.~Menache et al., ``Resource Management with Deep Reinforcement Learning,�\emph{ACM Workshop on Hot Topics in Networks (HotNets)}, Atlanta, USA, Nov. 09-10, 2016.
\bibitem{23}
N.~Liu, Z.~Li, Z.~Xu, J.~Xu, S.~Lin, Q.~Qiu, J.~Tang, Y.~Wang, ``A Hierarchical Framework of Cloud Resource Allocation and Power Management Using Deep Reinforcement Learning", \emph{IEEE International Conference on Distributed Computing Systems (ICDCS)}, Jun. 5-8 2017, pp. 372 - 382

\bibitem{24}
H.~Mao, R.~Netravali, M.~Alizadeh, ``Neural Adaptive Video Streaming with Pensieve," \emph{ACM SIGCOMM}, Aug. 21-25, 2017, pp.197-210

\bibitem{25}
Y.~Wei, F. Richard Yu, M.~Song, Z.~Han, ``User Scheduling and Resource Allocation in HetNets with Hybrid Energy Supply: An Actor-Critic Reinforcement Learning Approach," \emph{IEEE Trans. Wirel. Commun.}, vol.17, no.1, Jan. 2018, pp. 680-692.
\bibitem{26}
X.~Chen, L.~Jiao, W.~Li, X.~Fu, ``Efficient multi-user computation offloading for mobile-edge cloud computing," \emph{IEEE Trans. Netw.}, vol. 24, no. 5, pp. 2795-2808.
\bibitem{27}
T.~Liu, F.~Chen, Y.~Ma, Y.~Xie, ``An energy-efficient task scheduling for mobile devices based on cloud assistant," \emph{Future Generation Computer Systems}, vol.61, pp. 1-12

\bibitem{28}
Y.~B.~Lin, ``Reducing location update cost in a PCS network," \emph{IEEE Transactions on Networking}, vol. 5, no. 1, pp. 25-33, 1997.
\bibitem{29}
K.~Abboud,  W.~Zhuang, ``Stochastic Analysis of Single-Hop Communication Link in Vehicular Ad Hoc Networks," \emph{IEEE Trans. Vehicular Tech.}, vol. 65, no. 1, pp. 226-240, 2016
\bibitem{30}
J. Redmon, S. Divvala, R. Girshick, A. Farhadi, ``You only look once: Unified, real-time object detection,\emph{in proc IEEE Conference on Computer Vision and Pattern Recognition (CVPR)}," June 2016, pp. 779-788.



\bibitem{33}
H.~Chen, J.~Wang, Q.~Qi, H.~Sun, Y.~Li, ``Bilinear CNN Models for Food Recognition," \emph{International Conference on Digital Image Computing: Techniques and Applications}, 29 Nov.-1 Dec. 2017, pp. 1-6

\bibitem{35}
Y.~Zheng, Y.~Wang, L.~Zhang, J.~Wang, Q.~ Qi, ``A Tag-Based Integrated Diffusion Model for Personalized Location Recommendation," \emph{International Conference on Neural Information Processing}, 14-17 Nov. 2017, pp. 324-334


\end{thebibliography}
\end{document}